\documentclass[12pt,a4paper]{article}
  \usepackage{a4wide}
  \usepackage{latexsym}
  \usepackage{epsf}
  \usepackage{amssymb}
  \usepackage{graphicx}
  \usepackage{amsmath, cite}
  \usepackage{amsmath,amssymb,amsthm}
  \usepackage{verbatim}
  \usepackage{hyperref}
\renewcommand{\d}{\textrm{d}}

\date{\begin{flushleft}\today\end{flushleft}}

\newcommand{\be}{\begin{equation}}
\newcommand{\ee}{\end{equation}}

\begin{document}
\numberwithin{equation}{section}

\vspace{0.4cm}

\begin{flushright}
	UUITP - 22/21
\end{flushright}

\begin{center}

{\LARGE \bf{A higher-dimensional view \\ \vspace{0.4cm} on quantum cosmology}}

\vspace{2 cm} {\large U. H. Danielsson$^1$, D. Panizo$^1$,\\ \vspace{0.2cm} R. Tielemans$^2$, T. Van Riet$^{1,2}$}\\
 \vspace{1 cm} {\small\slshape $^1$ Institutionen f\"{o}r Fysik och Astronomi,\\ Box 803, SE-751 08 Uppsala, Sweden \\
 	$^2$ Instituut voor Theoretische Fysica, K.U.Leuven, \\Celestijnenlaan 200D, B-3001 Leuven, Belgium }
 
\vspace{.8cm} {\upshape\ttfamily ulf.danielsson, daniel.panizo \emph{at} physics.uu.se,\\ rob.tielemans, thomas.vanriet \emph{at} kuleuven.be }\\


\vspace{3cm}

{\bf Abstract} \end{center} {\small 
We argue that the choice of boundary condition for the wave function in quantum cosmology depends on the UV completion of general relativity. We provide an explicit example using a braneworld scenario in which a de Sitter cosmology is induced on the surface of a Coleman-de Luccia  bubble in a 5-dimensional AdS space. The corresponding boundary conditions are unambigously fixed by demanding consistency with the known physics of bubble nucleation and this selects the Vilenkin {  weighting} for the amplitude from a 4D viewpoint. }
\newpage
\section{Introduction}
Quantum cosmology (see e.g. \cite{Halliwell:1990uy}) aims at computing the wave function $\Psi$ to explain (features of) the universe. This is clearly an ambitious program and several aspects of its basic principles remain a source of debate in the literature. One particular issue concerns the choice of boundary conditions on the wave function. Just as in ordinary quantum mechanics, boundary conditions (and normalisation) are needed in order to uniquely fix solutions to Schr{\"o}dinger's equation. Such boundary conditions are often argued for by physical principles or by more formal principles such as the self-adjointness of the Hamiltonian. In quantum cosmology this is highly non-trivial to implement for several reasons. For starters, we do not know the full Hamiltonian since we lack the complete Hamiltonian of quantum gravity. The common attitude is to ignore the UV completion of gravity and to work in a semi-classical approximation, which is then further approximated by a mini-superspace approach, i.e., a drastic cut in the number of degrees of freedom. 

Given this, one can debate the choice of boundary conditions and reach various conclusions. Two natural choices are the Vilenkin choice (\emph{aka} the tunneling wave function) \cite{Vilenkin:1982de, Vilenkin:1984wp} and the Hartle-Hawking (HH) choice (\emph{aka} no-boundary proposal) \cite{Hartle:1983ai}. In mini-superspace, the difference can roughly be described as follows. At large scale-factor, the tunneling wave function can be seen as purely ``outgoing" waves, just as a wave function of a particle escaping a radioactive nucleus. The no-boundary proposal instead has fine-tuned ingoing and outgoing waves such that the wave function decreases towards the Big Bang singularity. More accurate descriptions are available, but we continue for now using this heuristic viewpoint.

The importance of the problem cannot be {  overstated} given the vastly different behaviors of the amplitudes of the two wave functions when considering their dependence on the cosmological constant (in Planck units)
\be\label{weightings}
\Psi_{\rm V} \approx e^{-c/\Lambda} \qquad \leftrightarrow \qquad \Psi_{\rm HH} \approx e^{c/\Lambda}\,,
\ee  
with $c$ a {  positive} numerical constant. The amplitude for the no-boundary proposal peaks at a small positive cosmological constant (cc) while the opposite is true for the tunneling wave function.  This would suggest that the most naive interpretations of the wave function might then be at odds with the tunneling wave function if one wants to address the cosmological constant problem (there is no real issue for inflation). But since we have not provided a full integration measure on the space of cosmological constants, such claims have no meaning.

{  Some claims, and counter-claims, appeared recently in \cite{Feldbrugge:2017kzv,Feldbrugge:2017fcc, DiazDorronsoro:2017hti, DiazDorronsoro:2018wro,Feldbrugge:2018gin, Vilenkin:2018dch,deAlwis:2018sec,Vilenkin:2018oja,DiTucci:2019dji}} concerning whether mathematical consistency requires one choice over the other, since certain saddle points dominate when choosing the correct integration contour. We find it more reasonable that the UV completion decides this by providing the extra physical input that informs us how the Hamiltonian at large curvatures really looks like. Imagine, for instance, the case discussed in \cite{Hertog:2021jyd} where the effective mini-superspace potential for the scale factor diverges to $ + \infty$ near zero scale-factor. This is like a hard wall model in quantum mechanics and the wave function should vanish at zero scale-factor, which would then select the no-boundary proposal. We refer to \cite{deAlwis:2018sec} for more arguments as to why physics rather than mathematics selects the boundary condition. 

{  One can also note recent claims that even the Hartle-Hawking saddle points potentially contribute
	a $e^{-c/\Lambda}$ amplitude instead of the inverse, see for instance \cite{DiTucci:2019dji}. We remain agnostic about these technical discussions and simply focus on the amplitude that is selected in our model. We will therefore refrain from using Hartle-Hawking or Vilenkin wave functions (boundary conditions), but instead will use the words {\it HH-weighting} and {\it Vilenkin-weighting}, keeping in mind their definition in equation \eqref{weightings}. In a follow-up work we will uncover more details of the wavefunction, but in this work we stick to a simple mini-superspace approach and focus purely on the amplitude.  }

We will show that we are able to address the choice of {  weighting} unambiguously in a model of cosmology that comes close to being UV completable within string theory \cite{Banerjee:2018qey}. One starts with a non-supersymmetric, perturbatively stable AdS$_5$ vacuum of string theory, which decays non-perturbatively by means of bubble nucleation leading to expanding thin wall bubbles containing the true vacuum. In string theory, such bubble walls are fundamental branes with matter and gauge forces localised on them. In \cite{Banerjee:2018qey, Banerjee:2019fzz, Banerjee:2020wov} it was demonstrated that a 4D de Sitter (dS) cosmology is induced on the bubble walls at late times. This is akin to the Karch-Randall mechanism \cite{Karch:2000ct} in some respects but different in others. In case of \cite{Karch:2000ct}, a closed dS universe would correspond to a bubble of AdS$_5$ identified across its boundary so that there is no outside. The bubbles we have in mind, expand into a pre-existing outside consisting of the false vacuum. As emphasised in \cite{Banerjee:2021yrb}, this scenario evades from the start the usual difficulties with dS model building in string theory \cite{Danielsson:2018ztv, Cicoli:2018kdo} -- including attempts to realize \cite{Karch:2000ct}. Instead of the kind of difficult uplifting needed in other constructions, the only thing needed is a brane that can mediate a non-perturbative decay of AdS$_5$.  Even more, this is perfectly in line with the Swampland ideas \cite{Ooguri:2016pdq, Brennan:2017rbf, Palti:2019pca} but uses them to go around the dS Swampland bounds.

\section{Hartle Hawking versus Vilenkin}
For the sake of simplicity, we consider cosmologies driven by a pure positive cosmological constant. Then the Friedmann equation for a 4D cosmology in case of positive spatial curvature becomes 
\be\label{Friedmann}
\dot{a}^2 = -1 + \frac{a^2}{R^2}\,,\qquad R^{-2} \equiv\frac{\kappa_4}{3}\rho_{\Lambda_4} = \Lambda_4 \,.
\ee
Here $R$ is the de Sitter radius and $\kappa_4= 8\pi G_4$.
The mini-superspace reduction of the Einstein-Hilbert action leading to the Friedmann equation is given by:
\be\label{action}
S = \frac{6\pi^2}{\kappa_4}\int \d \tau N \left(-\frac{a\dot{a}^2}{N^2} +a -  \frac{a^3}{R^2}\right)\,.
\ee
The lapse function $N$ from the FLRW metric
\be\label{eq:FLRW}
ds^2 = -N^2(\tau)\d \tau^2 + a(\tau)^2\d\Omega_3^2\,,
\ee
acts as a Lagrange multiplier whose constraint reproduces  (\ref{Friedmann}). 
From the action we find the canonical momentum $p=-\frac{12 \pi^2 a\dot{a}}{N}$. Quantizing using {  $p \rightarrow -i\frac{d}{da}$}, the Hamiltonian constraint becomes the Wheeler-deWitt (WdW) equation
\be\label{eq:WdW}
\frac{N}{a} \left(-\frac{1}{24 \pi^2}\frac{d^2}{da^2} + 6\pi^2 V(a)\right)\Psi(a) = 0\,,
\ee
where $V(a) =a^2-R^{-2}a^4$.

\begin{figure}[h!]
	\centering
	\includegraphics[width=8.6cm]{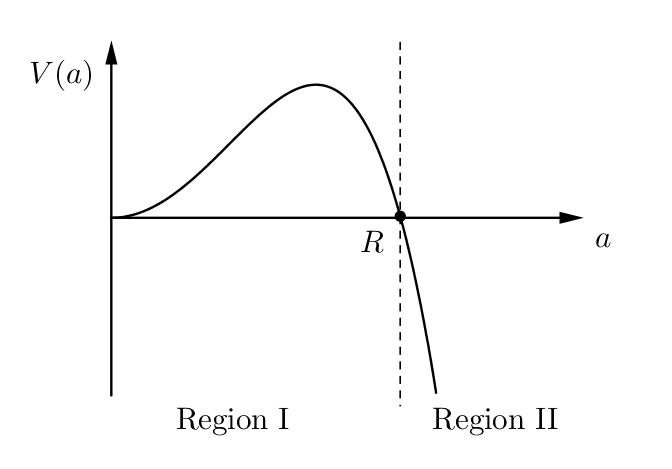}
	\caption{Effective potential.}
	\label{fig: potential}
\end{figure}

The plot  of the effective potential in Figure \ref{fig: potential} makes the analogy with tunneling through a barier manifest. 
There are two turning points: $a=0$ and $a=R$. The region between these two turning points is referred to as the Euclidean region and denoted region I in what follows, whereas the region $a>R$ is classical and denoted region II. 
The WKB solution is given by
\begin{align}
&\Psi_{\rm I}(a) = \frac{1}{|V(a)|^{1/4}}\left(ce^{S(a,0)}+ d e^{-S(a,0)}\right)\,,\\
&\Psi_{\rm II}(a) = \frac{1}{|V(a)|^{1/4}}\left(Ae^{iS(a,R)}+ B e^{-iS(a,R)}\right)\,,
\end{align}
where $c,d,A,B$ are complex constants that are related by the usual connection formulae and where we defined
\begin{equation}
S(a,a_i) \equiv \frac{12\pi^2}{\kappa_4}\int_{a_i}^a\sqrt{|V(a')|} da'\,.
\end{equation}

A choice $(c,d)$ or $(A,B)$ reflects a choice of boundary conditions. We also need a normalisation in order to fix them completely and it is common to take
\be
\lim\limits_{a\rightarrow 0}|V(a)|^{1/4}\Psi(a)=1\,.
\ee 
The Hartle-Hawking choice selects the growing exponential in region I by taking $(c,d)=(1,0)$:
\begin{equation}
\Psi_{\rm HH}(a) =\frac{1}{|V(a)|^{1/4}}\begin{cases} e^{S(a,0)}& \text{Region I}\\
2e^{S_0}\cos\left(S(a,R)-\frac{\pi}{4}\right) & \text{Region II}
\end{cases}.
\end{equation}
where 
\begin{equation}
S_0 \equiv S(R,0)=\frac{4\pi^2 R^2}{\kappa_4}\,.    
\end{equation}	
This leads to the following nucleation probability $P_{\rm HH}$:
\begin{equation} \label{PHH}
P_{\rm HH}\propto e^{2S_0}\,.
\end{equation}	
Vilenkin's choice is defined by selecting only outgoing waves in region II namely $(A,B) = (0,B)$:
\begin{equation}
\Psi_{\rm V}(a) \approx \frac{1}{|V(a)|^{1/4}}\begin{cases}
e^{S_0}e^{-S(a,0)+i\frac{\pi}{4}} & \text{Region I}\\
e^{-iS(a,R)} & \text{Region II}
\end{cases},
\end{equation}
where we approximated $c\approx 0$ since it is exponentially suppressed. 
We then find	\begin{equation}\label{Pvil}
P_{\rm V}\propto e^{-2S_0}\,.
\end{equation}
{  Our lightning overview of quantum cosmology has been very superficial but it suffices to convey the essential message of this paper in the next section; namely that a concrete UV completion will decide the boundary conditions near the ``Big Bang'' phase. A more detailed understanding of the quantum cosmology is of course attainable but beyond the scope of this letter.}

\section{Bubble universes}

Reference \cite{Banerjee:2018qey} proposed a new mechanism in string theory to obtain 4D de Sitter space. The main motivation behind this model is the persistent difficulties in string theory to construct stable and viable models with a positive cosmological constant \cite{Danielsson:2018ztv, Ooguri:2018wrx, Hebecker:2018vxz, Garg:2018reu}. Instead of compactifying six dimensions, a brane world scenario  {(for an overview see e.g. \cite{Maartens_2010} and references therein)} is considered that is similar in spirit to Karch-Randall \cite{Karch:2000ct} (or \cite{Hawking:2000kj}) but different in the crucial aspects. Instead of a bubble that is identified across its boundary so that it has no outside, the brane surface is a Coleman-de Luccia (CDL) bubble wall. The key insight is that the cosmology induced on a CDL bubble (mediating the decay of a non-supersymmetric 5D AdS vacuum) is 4D de Sitter space. The expansion of the bubble in 5D corresponds to the accelerated expansion of the 4D universe for an observed confined to the wall. 

This dark bubble scenario make use of key properties of string theory such as extra dimensions, branes, and the instability of non-SUSY AdS$_5$ vacua. Rather than being a problem, vacuum instabilities are turned into a virtue \cite{Banerjee:2021yrb}. A crucial advantage over previous brane-world scenarios is its natural UV embedding; the only requirement is an unstable 5D AdS vacuum whose primary decay proceeds through CDL bubbles, in line with the AdS weak gravity conjecture (WGC) \cite{Ooguri:2016pdq}. Another advantage is the clear physical picture and the ability to address issues in quantum cosmology as we now demonstrate.

In the notation of \cite{Banerjee:2018qey} we have that the cc's of the true and false vacuum are $\Lambda_\pm=-6k_\pm^2$ with $\Lambda_-<\Lambda_+<0$. The index $-$ ($+$) refers to quantities evaluated in the true (false) vacuum. The expansion of the bubble is governed by the Israel junction conditions:
\be 
\label{juncti}
\sigma=\frac{3}{\kappa_5}\left(\sqrt{k^2_{-}+\frac{1+\dot{a}^2}{a^2}}-\sqrt{k_{+}^2+\frac{1+\dot{a}^2}{a^2}}\right)\,,
\ee
where $\sigma$ is the tension of the brane, and $\kappa_5= 8\pi G_5$. The induced metric on the bubble is simply given by (\ref{eq:FLRW}).
Properly squaring the junction condition generates a 4D Friedmann equation with positive cosmological constant (\ref{Friedmann}). To achieve this, one needs to identify
\be 
\label{5to4}
\kappa_4=\frac{2k_- k_+ }{k_- -k_+} \kappa_5\,.
\ee
From here onwards we work in the physical approximation of large $k_{\pm}$ in which the 4D vacuum energy simplifies to \footnote{Squaring the junction condition also yields an exact value for the 4D vacuum energy, namely 
	$		\rho_{\Lambda_4} = \frac{1}{\kappa_4}\left(\frac{9}{4\kappa_5\sigma^2}\left(\frac{9}{\kappa_5^2}\left(k_-^2-k_+^2\right)-\sigma^2\right)^2 - k_+^2\right).$ Expanding this expression close to the critical tension, (\ref{lambdax}) is obtained. }:
\be 
\label{lambdax}
\rho_{\Lambda_4} \equiv \frac{3}{\kappa_4} R^{-2} = \frac{3\left(k_{-}-k_{+}\right)}{\kappa_5}- \sigma.  
\ee
A bubble can only nucleate if the tension is less than a critical value $\sigma_{cr} = \frac{3}{k_{5}}(k_{-}-k_{+})$ , and this condition automatically leads to $\rho_{\Lambda_4}>0$. 

In \cite{Banerjee:2019fzz, Banerjee:2020wix, Banerjee:2020wov} this model was studied in more detail introducing radiation using a bulk black hole and dust from a cloud of strings in the bulk. Using the Gauss-Codazzi equations, it was also shown that the full relativistic equations of 4D general relativity are recovered.

\section{Quantum mechanical description}

We now study the quantum nucleation of such a bubble by closely following the treatment of \cite{Ansoldi:1997hz} for bubble nucleation in 4D and extend it to 5D. The 5D action is given by
\be
\begin{aligned}
	S=&\frac{1}{2\kappa_5} \int d^5 x \sqrt{|g|} \left( R^{(5)} - 2 \Lambda \right) -\sigma \int d^4 \zeta \sqrt{|\eta|} \\ & +\frac{1}{\kappa_5} \oint d^4 x \sqrt{|h|} K\,. \label{5Daction}
\end{aligned}
\ee
The second term describes the brane-shell with tension $\sigma$ and induced metric $\eta$ with brane-coordinates $\zeta$.  The 5D metric is given by
\begin{equation}
\d s_{\pm}^2 = -A_{\pm}(r)\d t_{\pm}^2+\frac{\d r^2}{A_{\pm}(r)} + r^2\d\Omega_{3}^2, \label{eq: metric int/ext}
\end{equation}
where $A_{\pm}(r) = 1 -\frac{\Lambda_{\pm}}{6}r^2$.

The shell glues the two spacetimes together at a radial coordinate $r=a(\tau)$ and its metric coincides with (\ref{eq:FLRW}).
From \eqref{eq: metric int/ext} and \eqref{eq:FLRW} one can then deduce that
\begin{equation}
\dot{t}_{\pm}=\frac{\d t_{\pm}}{\d \tau} = \frac{\beta_{\pm}}{A_{\pm}}\,,\qquad \beta_{\pm} = \sqrt{A_{\pm}N^2+\dot{a}^2}\,.
\end{equation}

The on-shell action receives three contributions: the bulk piece, the shell contribution and the boundary term.
Summing all terms, and ignoring terms that do not affect the dynamics of the shell, we find the mini-superspace Lagrangian to be
\be
L=\frac{6\pi^2}{\kappa_5} \left[-a^2 \dot{a} \tanh^{-1} \frac{\dot{a}}{\beta} + a^2 \beta \right]_+^- -2\pi^2 a^3 \sigma N\,.
\ee
Expanding to quadratic order in $\dot{a}$, using (\ref{5to4}), equation (\ref{action}) can be recovered. Using this we can go ahead and study nucleation, where $R$ will be the radius of the nucleated bubble. 

We now argue that the physics of bubble nucleation in 5D then fixes the {  amplitude} to be of the tunneling type. Our approach is simple: we verify that Vilenkin's tunneling amplitude exactly matches the known CDL amplitudes, as expected from the physical picture of tunneling.

There are a few different ways to calculate the nucleation probability $P=e^{-B}$ that all yield the same result.
Using the WKB wall penetration probability we have $P=e^{-B}$ with $B=2 \int d\tau\:p$ ($p$ is the canonical momentum) and we recover Vilenkins choice (\ref{Pvil})
\be\label{Vil}
B= \frac{24\pi^2}{\kappa_4} \int_0^R da \sqrt{a^2-\frac{a^4}{R^2}} =\frac{8\pi^2 R^2}{ \kappa_4}\,.
\ee

One can also use the approach of Brown and Teitelboim \cite{Brown:1988kg}. The Euclidean instanton is then obtained by integrating  (\ref{5Daction}) over a $O(5)$ symmetric ball of radius $R$, with a 4D sphere as boundary and corresponds to a bounce. Evaluating expression (6.4) in \cite{Brown:1988kg} with $R^{(5)}_\pm=-20k^2_\pm$, which follows from the equations of motion, we find:
\be
B= \sigma A_4 + \frac{1}{\kappa_5}\left[ 4 k^2 V_5 \left(R, k\right) -\frac{4}{R} \beta A_4 \right]_+^-\,,
\ee
with $A_4=8\pi^2 R^4/3$, and $dV/dR=A_4/\beta$. Extremizing using $dB/dR=0$ implies the junction condition and fixes $R$ to the critical value. Expanding in large $k$, we have
\be
V_5=\frac{A_4}{4k}\left(1-\frac{1}{k^2 R^2} \right)\,.
\ee 
Inserting this, and replacing $\kappa_5$ by $\kappa_4$ using (\ref{5to4}), we recover (\ref{Vil}).

Another nice feature of our 5D UV completion is that we have control over an infinite tower of corrections to the 4D quantum cosmological model.  Going back to the full expression, the canonical momentum is given by
\be
\cosh \left( \frac{\kappa_5 p}{6\pi^2 a^2}\right) = \frac{\beta_- \beta _+ - \dot{a}^2}{N^2\sqrt{A_- A_+}}\,,
\ee
and the Hamiltonian plays the role of a constraint imposing the junction condition:
\be
H=2\pi^2 N a^3 \left( \sigma-\frac{3\left(\beta_- -\beta_+ \right)}{a\kappa_5} \right) =0\,.
\ee
Expressed in terms of the canonical momentum the Hamiltonian constraint becomes
\be
\begin{aligned}
	H= & - \frac{6\pi^2}{\kappa_5} \left( A_- + A_+-2\sqrt{A_- A_+} \cosh  \left( \frac{\kappa_5 p}{6\pi^2 a^2}\right) \right)^{1/2} \\ & + 2\pi^2 N a^3 \sigma =0\,.
\end{aligned}
\ee
We can quantize the system to obtain the WdW-equation by making the replacement
\be
p \rightarrow - \frac{i }{a^{3/2}}\frac{d}{da} a^{3/2} .
\ee
For general $p$ the equation is of infinite order in $p$ and turns into a difference equation. We focus on the limit of small $p$, where the Hamiltonian becomes quadratic in $p$. This is the limit that is relevant for the case of a small cosmological constant compared to fundamental scales. In this limit, we recover (\ref{eq:WdW})
\be
\left(-\frac{1}{24 \pi^2}\frac{1}{a^{3/2}}\frac{d^2}{d a^2}( a^{3/2} \psi) + 6\pi^2 V(a)\right)\psi = 0\,,
\ee
with a different normalization of the wave function that is easily understood. Here, the wave function $\psi$ is supported in four spatial dimensions, and is related to the wave function in minisuperspace throught $\Psi= a^{3/2} \psi$. Note that $\int a^3\lvert \psi \rvert ^2 = \int \lvert \Psi \rvert ^2$  is used for normalization.

\section{Conclusion}

We conclude that {  Vilenkin's amplitude} in quantum cosmology can be understood as the nucleation probability of a bubble of true vacuum in an unstable AdS$_5$ space. Our understanding of the physics of CDL bubbles (see \cite{Ghosh:2021lua})  translates to an understanding of more involved issues in quantum cosmology such as the choice of boundary conditions{ , which affect the amplitudes}.

In this model there is no Big-Bang singularity to worry about and all physics at all length scales that are involved in the process are essentially understood. It is semi-classical gravity in 5D, which can be embedded in string theory and provides our UV completion of 4D cosmology. Indeed, the Big Bang singularity is not present in 5D and would correspond to the fiducial zero size of the bubble, which is not a physical solution to worry about. The undetermined coefficients in the mini-superspace wave function are then completely fixed, since it has to be consistent with the standard story of thin wall CDL tunneling (or the particular Brown-Teitelboim  incarnation in this case). {  From the point of view of a 4D observer, the nucleation of the closed bubble universe is a creation out of nothing. From a 5D point of view, one still needs to explain the origin of the AdS space time. This question we do not address and leave for further string theoretic studies, possibly within the framework of a multiverse.}

{  In \cite{Feldbrugge:2017kzv,Feldbrugge:2017fcc, DiazDorronsoro:2017hti, DiazDorronsoro:2018wro,Feldbrugge:2018gin, Vilenkin:2018dch,deAlwis:2018sec,Vilenkin:2018oja,DiTucci:2019dji} it has been debated whether quantum cosmology is potentially unstable against fluctuations. We believe that the arguments in \cite{Vilenkin:2018dch,deAlwis:2018sec,Vilenkin:2018oja, DiTucci:2019dji} settles this, and take away worries about such instabilities in the case of Vilenkin weighting. The bottom line is that the path integral needs to be supplied by extra terms to properly take into account boundary conditions. These terms contribute to the amplitude and can be interpreted as appropriate wave functions for the fields involved. Our interpretation of the Vilenkin tunneling proposal in terms of the nucleation of a higher dimensional bubble, intuitively explains why instabilities are not to be expected. As argued in \cite{deAlwis:2018sec}, the path integral is just a way to solve the WdW equation. Examining solutions of the latter is the simplest way to sort out the physics of the problem. }

It is useful to compare our results with the model of Karch-Randall \cite{Karch:2000ct}. There, a closed dS universe is represented by a bubble with its inside identified with itself across its boundary. Such a bubble has no outside and cannot nucleate in to a pre-existing space time. Interestingly, \cite{Hawking:2000kj} concluded that the quantum creation of such a universe would be described by the {  Hartle-Hawking amplitude}. This is in contrast with the nucleating bubbles we have studied in this paper, which tunnel into existence as a false vacuum decays and thus need to be described by Vilenkin's tunneling {  amplitude}. The swampland conjectures suggest that it is the latter possibility that has a chance of being realized in string theory (see \cite{Basile:2020mpt} for a concrete suggestion).

\section*{Acknowledgments}
We like to thank Alex Vilenkin for comments on an earlier draft. The work of RT and TVR is supported by the KU Leuven C1 grant ZKD1118C16/16/005. TVR is furthermore supported by the FWO fellowship for sabbatical research and would like to thank Uppsala University for hospitality. DP would like to thank the Centre for Interdisciplinary Mathematics (CIM) for financial support.

\small{
\bibliography{refs}}
\bibliographystyle{utphys}

\end{document}